# Strong correlation behavior and Strong coupling superconductivity in $(Ti_{1/4}Hf_{1/4}Nb_{1/4}Ta_{1/4})_{1-x}Ni_x$ with the rich magnetic element Ni

*Zijun Huang[1,#], Tong Li[1,#], Longfu Li[1,#], Rui Chen[1], Zaichen Xiang[1], Shuangyue Wang[1], Jingjun Qin[1], Yucheng Li[1], Lingyong Zeng [1,5,\*], Dinghua Bao[1], Huixia Luo [1,2,3,4\*]*

[1] School of Materials Science and Engineering, Sun Yat-sen University, Guangzhou 510006, China

[2] State Key Laboratory of Optoelectronic Materials and Technologies, Sun Yat-sen University, Guangzhou 510006, China

[3] Guang-dong Provincial Key Laboratory of Magnetoelectric Physics and Devices, Sun Yat-sen University, Guangzhou 510006, China

[4] Key Lab of Polymer Composite & Functional Materials, Sun Yat-sen University, Guangzhou 510006, China

[5] Device Physics of Complex Materials, Zernike Institute for Advanced Materials, University of Groningen, 9747 AG Groningen, the Netherlands

[#] Z. Huang, T. Li, and L. Li contributed equally to this work.
\*Corresponding author E-mail: *l.zeng@rug.nl* or *zengly57@mail.sysu.edu.cn* (L. Zeng); *luohx7@mail.sysu.edu.cn* (H. Luo)

**Abstract**

Searching for new superconductors, especially unconventional superconductors, has been studied extensively for decades but remains one of the major outstanding challenges in condensed matter physics. Medium/high-entropy alloys (MEAs-HEAs) are new fertile soils of unconventional superconductors and generate widespread interest and questions on the existence of superconductivity in highly disordered materials. Here, we report on the effect of Ni-doped on the crystal structure and superconductivity properties of strongly coupled TiHfNbTa MEA. XRD results indicate that the maximum solid solution of $(Ti_{1/4}Hf_{1/4}Nb_{1/4}Ta_{1/4})_{1-x}Ni_x$ is about 7.7%. Resistivity, magnetic susceptibility, and specific heat measurements demonstrated that $(Ti_{1/4}Hf_{1/4}Nb_{1/4}Ta_{1/4})_{1-x}Ni_x$ HEAs are all bulk type-II superconductors and follow the trend of the increase of $T_c$ with the increase of Ni-doped contents. The specific heat jump $(\Delta C_{el}/\gamma T_c^{mid})$ of all $(Ti_{1/4}Hf_{1/4}Nb_{1/4}Ta_{1/4})_{1-x}Ni_x$ are much larger than the BCS value of 1.43, suggesting all these HEAs are strongly coupled superconductors. Additionally, large Kadawaki-Woods ratio (KWR) values suggest that there is a strong electron correlation effect in this system. The $(Ti_{1/4}Hf_{1/4}Nb_{1/4}Ta_{1/4})_{1-x}Ni_x$ HEA system is a new ideal material platform for the study of strong correlation behavior and strongly coupled superconductivity, which provides an insight into the physics of high-temperature superconductors or other unconventional superconductors.

# I. Introduction

Medium- and high-entropy alloys (MEAs-HEAs) are metallic materials studied widely in recent years. Different from conventional alloys, which often have one dominant element, MEAs-HEAs usually comprise various kinds of principal metallic elements in near-equimolar ratios [1-3]. In general, MEAs-HEAs tend to form disordered solid solutions in simple structures, such as body-centered cubic (BCC), face-centered cubic (FCC), or hexagonal closed-pack (HCP), and elements occupy a lattice site randomly [1-5], which brings MEAs-HEAs very high mix entropy and make MEAs-HEAs steady easily in simple structures. Generally speaking, MEAs' mix entropy $\Delta S_{mix}$ ranges between 0.69R and 1.60R, while HEAs' $\Delta S_{mix}$ can reach 1.6R and even higher [6]. The great atomic disorder not only brings MEAs-HEAs stable structures but also produces an intense lattice distortion effect, which endows MEAs-HEAs with lots of excellent properties, including high hardness [7], high specific strength [8,9], fracture-wear-corrosion-oxidation resistances [10] and so on. There is no doubt that MEAs-HEAs can be brand new kinds of functional materials and their outstanding performance makes them promising candidates for novel applications.

Besides exceptional mechanical properties, some MEAs-HEAs also exhibit superconductivity, which was first found in the Nb-Ta-Ti-Zr-Hf system in 2014 [11]. Since then, the study on MEA-HEA superconductors has begun quickly and then been in full swing, as well as aroused questions on the existence of superconductivity in such highly disordered materials [12-15]. Because Cooper pairs are generally difficult to form in such chaotic conditions, the superconductivity of MEAs-HEAs is so remarkable that it may embody a new kind of superconductivity mechanism. Meanwhile, the critical temperatures ($T_c$s) of HEA superconductors show robustness against both disorder and high pressure [14,15], and display a unique kind of dependence on valence electron count (VEC) [14], which differs from those of amorphous and crystalline alloy superconductors. When we take BCC-type HEA superconductors' $T_c$ as a function of VEC, the highest $T_c$ is reached at a VEC of about 4.7 [14], which matches the Matthias rules for transition metal superconductors [16]. Also, it is important to note that the Ta-Nb-Hf-Zr-Ti HEA superconductor originates from Nb-Ti-based binary alloys, which are currently the most widely utilized materials for superconducting magnets [17]. So, combining 3d, 4d, and 5d elements like Ti, Hf, Nb, and Ta may be the feasible way to discover new MEAs-HEAs. Recently, a Nb-Ti-based MEA superconductor TiHfNbTa

(VEC = 4.6) shows extremely strong coupling *s*-wave superconductivity [18], which is uncommon due to its usual occurrence in cuprates, pnictides, and other unconventional superconductors. On the other hand, apart from MEAs-HEAs based on non-magnetic transition metals, magnetic Cr/Co/Ni-based MEAs-HEAs show rich physical/chemical properties [17,19]. Among these elements, Ni has an important effect on HEAs' mechanical properties like ductility [20], corrosion resistance [21], and hardness [22]. However, there are few studies on magnetic element Ni doping in the superconducting MEAs-HEAs. Whether the incorporation of magnetic element Ni can confer unique physical properties on superconducting MEAs-HEAs, especially the strong coupling superconducting TiHfNbTa [18]. Diving into these questions is beneficial for us to design new functional materials and reveal the unique superconducting mechanism behind the MEAs-HEAs further. Therefore, based on our previous findings, we aim to explore the effect of Ni-doped content on the structure and superconductivity of TiHfNbTa.

Here, we report a series of $(Ti_{1/4}Hf_{1/4}Nb_{1/4}Ta_{1/4})_{1-x}Ni_x$ ($0 < x \leq 0.077$) MEAs-HEAs, which are prepared by an arc-melted method. A detailed investigation has been conducted on the relationship between Ni content and superconducting properties. X-ray diffraction (XRD) results confirm that all these $(Ti_{1/4}Hf_{1/4}Nb_{1/4}Ta_{1/4})_{1-x}Ni_x$ ($0 < x \leq 0.077$) MEA-HEA compositions adopt a BCC structure. Temperature-dependent resistivity and magnetic susceptibility results indicate that $(Ti_{1/4}Hf_{1/4}Nb_{1/4}Ta_{1/4})_{1-x}Ni_x$ ($0 < x \leq 0.077$) MEAs-HEAs are all type-II superconductors and follow the trends: $T_c$ increases with the increase of Ni-doping content, and the coupling parameter $\Delta C_{el}/\gamma T_c$ changes from 2.89 to 2.44, all above the BCS weak coupling limit 1.43. It can be concluded that the doping of Ni may weaken the electron-phonon coupling slightly, although all of them represent the behavior of strong electron-phonon coupling. Our work explains to some extent what role Ni plays in the superconducting properties of MEA-HEA materials.

## II. Experimental Details

The polycrystalline MEA-HEA samples $(Ti_{1/4}Hf_{1/4}Nb_{1/4}Ta_{1/4})_{1-x}Ni_x$ ($0 < x \leq 0.077$) were synthesized by a conventional arc-melting method. Stoichiometric amounts of high-purity metal elements powder were weighed and pressed into pellets after mixing completely. Subsequently, the pellets were put into a copper furnace to perform arc-melting under an argon atmosphere. To ensure the phase homogeneity of samples, we flipped and remelted them 3 times. The structural information of these samples was obtained by X-ray diffraction (XRD) using the MiniFlex of Rigaku (Cu K$\alpha$ radiation). The XRD patterns were collected from 20° to 100° by a scan speed of 1° /min. Secondary electron images, backscattered electron images and elemental analysis were examined by scanning electron microscopy (SEM) coupled with energy-dispersive X-ray spectroscopy (EVO MA10, CARL ZEISS). The accelerting voltage of electron beam for both secondary electron images and backscattered electron images were set at 20kV. In addition, the electrical, magnetism, and heat capacity performance were measured by a Physical Property Measurement System (PPMS, Quantum Design).

## III. Results and Discussion

The bulk XRD patterns of polycrystalline HEA $(Ti_{1/4}Hf_{1/4}Nb_{1/4}Ta_{1/4})_{1-x}Ni_x$ ($0 < x \leq 0.077$) are presented in Figure 1(a), indicating that these samples are pure single phases without any impurity. These XRD patterns can be indexed in the BCC structure with the $Im$-$3m$ space group, revealing that the doping of Ni does not change the crystal structure of the TiHfNbTa matrix. In addition, the diffraction peaks are broadened, owing to the disorder in the microscopic scale. Detailed diffraction data within the 2θ range of 37° to 38.5° is shown on the right-hand side of Figure 1(a), which is the diffraction peaks of the plane (110). These peaks shift to a higher angle with the increase in Ni content, which means that the lattice constant $a$ decreases with the increase in Ni content. This behavior originates from Ni's relatively small atomic radius (124 pm) compared to Ti (147 pm), Nb (146 pm), Hf (159 pm), and Ta (146 pm). Rietveld refinement was used to determine the lattice constant $a$ of these samples. The relationship between the lattice parameter, unit cell volume, and content $x$ is demonstrated in Figure 1(b). The atomic occupation of the lattice site is unrestricted by atomic species, which indicates the microscopic disorder of the series of samples. SEM-EDX was applied to the series of samples to confirm the microscopic uniformity. Figure

1(c) displays the SEM, BSE images, and elemental mappings of $(Ti_{1/4}Hf_{1/4}Nb_{1/4}Ta_{1/4})_{1-x}Ni_x$ ($0 < x \leq 0.077$) HEA, which demonstrates that the series of samples are microscopically homogeneous. The elemental distribution is close to the designed values within the allowable margin of error, which is demonstrated in Table S1. In addition, EDX line scans and point analysis were applied to confirm the microscopic uniformity of the series of samples. Figure S1 displayed the data of EDX line scans and point analysis. The data of the elements ratio of point/line analysis have been sorted out in Table S2 and Table S3, illustrating the microscopic homogeneity of these samples.

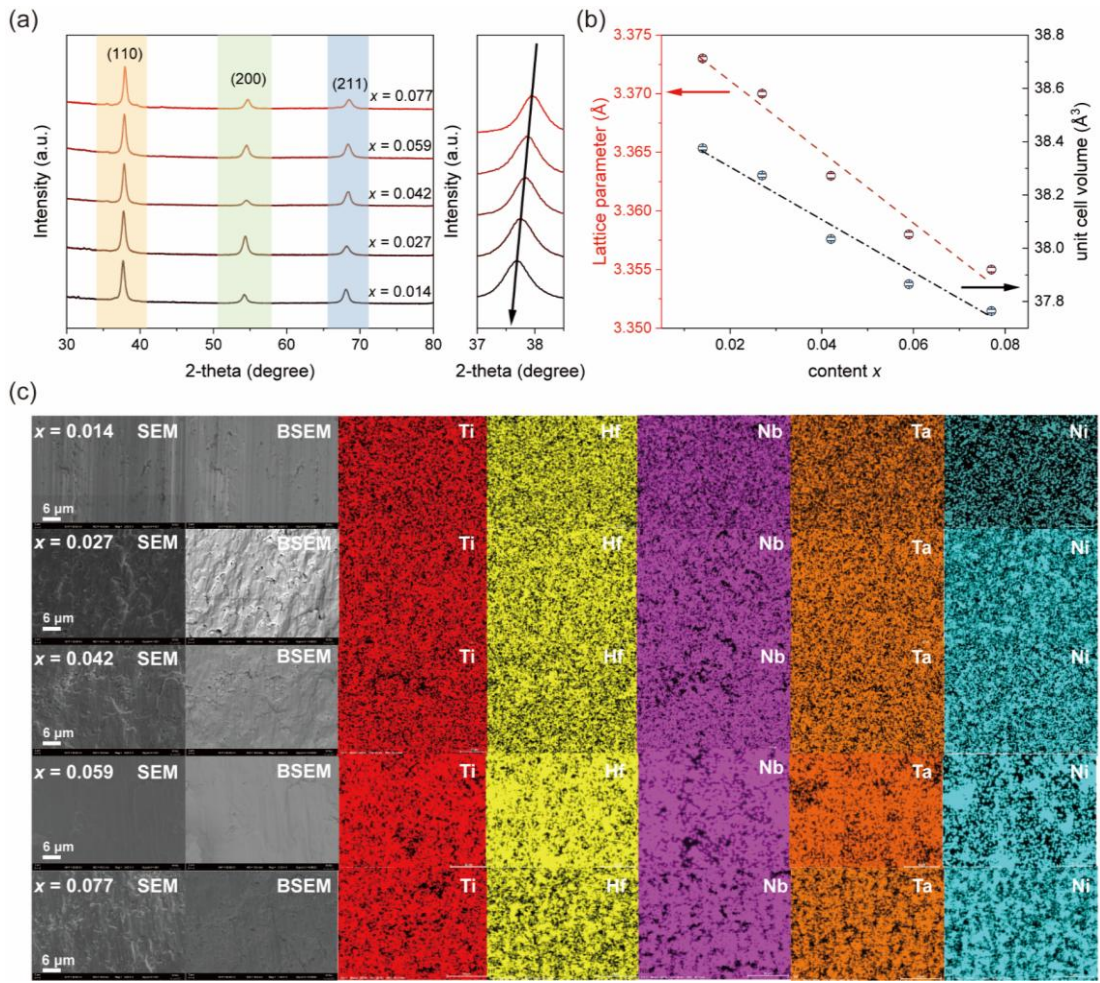

**Figure 1.** (a) XRD patterns of the $(Ti_{1/4}Hf_{1/4}Nb_{1/4}Ta_{1/4})_{1-x}Ni_x$ ($0 < x \leq 0.077$). The (hkl) Miller indices denote the reflections from the BCC structure and diffraction related to the planes (110) in detail. (b) The relationship between the lattice parameter, unit cell volume, and content $x$. (c) SEM, BSEM imgaes and EDX elemental mappings of the $(Ti_{1/4}Hf_{1/4}Nb_{1/4}Ta_{1/4})_{1-x}Ni_x$ ($0 < x \leq 0.077$) HEA.

To investigate the electrical transport properties of HEA $(Ti_{1/4}Hf_{1/4}Nb_{1/4}Ta_{1/4})_{1-x}Ni_x$ ($0 < x \leq 0.077$), resistivity measurements were systematically conducted. Figure 2(a) displays the temperature-dependent resistivity between 300 and 1.8 K in the zero field. This series of samples exhibit metallic behavior, in which the resistivity decreases slowly as the temperature decreases. We calculated the residual resistivity ratio (RRR), which is close to 1 (shown in Table 1). The resistivity drops sharply to zero at low temperatures, revealing that there exists superconductivity. To characterize the superconductivity, Figure 2(b) shows the temperature-dependent resistivity between 6 and 8 K. The $T_c$ is defined as the midpoint of the temperature from the normal state value to the zero-resistivity value. The $T_c$ rises from 6.55 K to 7.35 K monotonically as the Ni content increases, implying that the doping with magnetic elements Ni elevates the $T_c$s instead of suppressing the superconductivity.

To estimate the upper critical fields of HEA $(Ti_{1/4}Hf_{1/4}Nb_{1/4}Ta_{1/4})_{1-x}Ni_x$ ($0 < x \leq 0.077$), we conducted resistivity measures under different fields. Figure 2(c) and Figure S1 show the magneto-transport behavior of the $(Ti_{1/4}Hf_{1/4}Nb_{1/4}Ta_{1/4})_{1-x}Ni_x$ ($0 < x \leq 0.077$) samples. The $T_c$ decreases with the enhancement of perpendicular fields. The upper critical fields at 0 K $\mu_0H_{c2}(0)$ were determined by extrapolating the temperature-dependent data using the Ginzburg-Landau (G-L) formula: $\mu_0H_{c2}(T) = \mu_0H_{c2}(0) \times \frac{1-(T/T_c)^2}{1+(T/T_c)^2}$ [23]. Although G-L theory is especially applicable near $T_c$, the above equation has been proved to be satisfied in a much wider temperature regime [24]. Figure 2(d) displays the fitting curve of the series of samples. Besides, the upper critical field is also estimated by the Werthamer–Helfand–Hohenberg (WHH) formula: $H_{c2}(0) = -0.693T_c(\frac{dH_{c2}}{dT})_{T=T_c}$ [25]. The $\mu_0H_{c2}(0)^{WHH}$ follows a similar trend as the $\mu_0H_{c2}(0)^{GL}$. Generally, the $\mu_0H_{c2}(0)$ increases with the Ni content $x$. Some research in MEA/HEA superconductors found that there is a positive correlation between the upper critical field and the entropy, the upper critical field appears to increase with increasing entropy [4,14,26]. At the same time, the upper critical field is also proportional to $\gamma\rho_N$ in the dirty limit, where $\rho_N$ is the resistivity just above $T_c$ in the normal state [27]. However, the $\gamma\rho_N$ of our samples are non-monotonic, this may be the reason for the non-monotonic behavior of $\mu_0H_{c2}(0)$ as a function of $x$. The Pauli limiting field $H^p$ can be obtained by this equation: $H^p = 1.86 \times T_c$. The $\mu_0H_{c2}(0)$ for content $x = 0.077$ is 11.26 T which is smaller than the $H^p = 13.69$ T. All the values are summarized in Table 1.

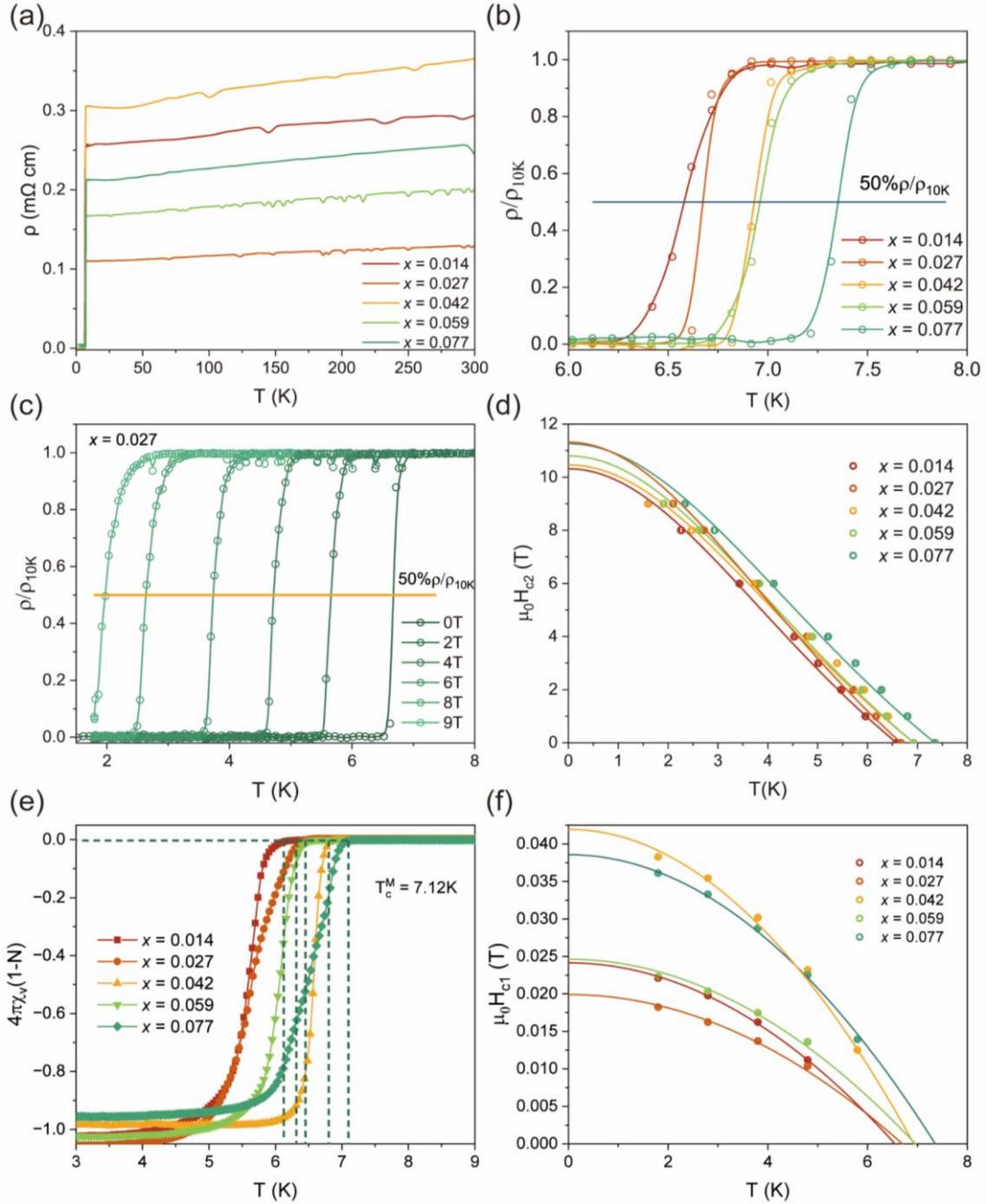

**Figure 2.** (a) Temperature-dependent resistivity of $(Ti_{1/4}Hf_{1/4}Nb_{1/4}Ta_{1/4})_{1-x}Ni_x$ ($0 < x \leq 0.077$). (b) The temperature-dependent resistivity at low temperatures. (c) The resistivity transition at the field of 0 - 9 T for $x = 0.027$. (d) The temperature-dependent upper critical field $\mu_0H_{c2}$ with the Ginzburg-Landau function fitting. (e) The temperature-dependent magnetic susceptibility for $(Ti_{1/4}Hf_{1/4}Nb_{1/4}Ta_{1/4})_{1-x}Ni_x$ ($0 < x \leq 0.077$). (f) The temperature-dependent lower critical fields for $(Ti_{1/4}Hf_{1/4}Nb_{1/4}Ta_{1/4})_{1-x}Ni_x$ ($0 < x \leq 0.077$).

Table 1 Superconducting parameters for the $(Ti_{1/4}Hf_{1/4}Nb_{1/4}Ta_{1/4})_{1-x}Ni_x$ $(0 < x \leq 0.077)$.

| x content | 0 [18] | 0.014 | 0.027 | 0.042 | 0.059 | 0.077 |
|---|---|---|---|---|---|---|
| $\Delta S_{mix}$ | 1.39 | 1.44 | 1.47 | 1.50 | 1.53 | 1.55 |
| $T_c$ (K) | 6.75 | 6.55 | 6.70 | 6.93 | 6.96 | 7.36 |
| RRR | - | 1.15 | 1.18 | 1.20 | 1.21 | 1.15 |
| $\mu_0 H_{c1}$ (T) | 0.046 | 0.024 | 0.020 | 0.042 | 0.024 | 0.039 |
| $\mu_0 H_{c2}$ (T)$^{GL}$ | 10.46 | 10.32 | 11.32 | 10.46 | 10.80 | 11.26 |
| $\mu_0 H_{c2}$ (T)$^{WHH}$ | 8.64 | 8.39 | 9.26 | 9.01 | 9.07 | 9.18 |
| $\gamma$ (mJ mol$^{-1}$ K$^{-2}$) | 4.70 | 4.62 | 4.66 | 4.20 | 4.71 | 4.30 |
| $\beta$ (mJ mol$^{-1}$ K$^{-4}$) | 0.24 | 0.23 | 0.18 | 0.22 | 0.21 | 0.21 |
| $\Delta C_{el}/\gamma T_c$ | 2.88 | 2.89 | 2.79 | 2.77 | 2.56 | 2.44 |
| $\Theta_D$ (K) | 199.90 | 203.57 | 220.00 | 206.17 | 211.15 | 211.31 |
| $2\Delta_0/k_B T_c$ | 5.02 | 5.01 | 4.92 | 4.91 | 4.71 | 4.60 |
| $\omega_{ln}$ (K) | 38.00 | 39.26 | 44.74 | 47.32 | 57.24 | 67.31 |
| $\lambda_{ep}$ | 0.83 | 0.81 | 0.79 | 0.83 | 0.82 | 0.84 |
| VEC | 4.50 | 4.57 | 4.65 | 4.73 | 4.82 | 4.92 |

To confirm the bulk superconductivity of the series of samples, we conducted the magnetic-susceptibility measurements under a magnetic field of 3 mT. Figure 2(e) shows the temperature-dependent magnetization curves of $(Ti_{1/4}Hf_{1/4}Nb_{1/4}Ta_{1/4})_{1-x}Ni_x$ ($0 < x \leq 0.077$) HEAs, which demonstrates the diamagnetism of the series of samples. All samples show a strong diamagnetic signal below their $T_c$s. Demagnetization corrections were applied to the magnetization data using the demagnetizing factor N. The corrected susceptibility $4\pi\chi_v(1-N)$ attains the theoretical value of -1 below $T_c$s, demonstrating a 100 % Meissner volume fraction, confirming the bulk superconductivity of the series of samples. The $T_c$ obtained from the magnetization measurement was close to the resistivity measurement, indicating the high quality of the series of samples. To calculate the lower critical fields $\mu_0 H_{c1}$ of the series of samples at 0 K, the field-dependent magnetization measurement was conducted. Figure S2 presents the magnetization isothermal curves, revealing a systematic reduction of the $\mu_0 H_{c1}$ with increasing temperature. The $\mu_0 H_{c1}$ can be obtained by Figure S2, which also displays the relationship between H and M-$M_{fit}$. In addition, we calculate the lower critical fields at 0K in Figure 2(f) by using the formula $\mu_0 H_{c1}(T) = \mu_0 H_{c1}(0) \times (1 - (T/T_c)^2)$. To analyze the superconductivity transition of HEA $(Ti_{1/4}Hf_{1/4}Nb_{1/4}Ta_{1/4})_{1-x}Ni_x$ ($0 < x \leq 0.077$), ac-magnetic susceptibility was conducted. Figure S4 displayed the real and imaginary part of AC magnetic susceptibility of $(Ti_{1/4}Hf_{1/4}Nb_{1/4}Ta_{1/4})_{1-x}Ni_x$ ($x$ = 0.027, 0.042, 0.077)

HEAs. The χ' drop sharply at $T_c$s, exhibiting diamagnetic behavior. At the same time, the χ" exhibits a single peak below the critical temperature, which is defined as the onset point of the magnetic susceptibility beginning to deviate from normal-state behavior, indicating significant hysteresis loss in the superconducting state. The data of AC magnetic susceptibility confirms the existence of single superconductivity transition.

Heat capacity measurements under zero field were applied to the HEA $(Ti_{1/4}Hf_{1/4}Nb_{1/4}Ta_{1/4})_{1-x}Ni_x$ ($0 < x \leq 0.077$) to take further insight into the electron-phonon coupling. As demonstrated in Figure 5, the normal-state specific heat data can be modeled by the expression: $C_p/T = \gamma + \beta T^2$ where $\gamma$ denotes the Sommerfeld constant reflecting the electronic contribution to the specific heat, and the $\beta T^2$ term represents the lattice contribution. The fit gives $\gamma = 4.297$ mJ mol$^{-1}$ K$^{-2}$ and $\beta = 0.206$ mJ mol$^{-1}$ K$^{-4}$ for $x = 0.077$. In low temperatures, the specific heat jumps sharply at $T_c$s, indicating a superconductivity transition. $T_c$s increase with the enhancement of content $x$, which is consistent with the result of resistant measurement. To investigate the behavior of electrons in low temperatures, the temperature-dependence specific heat of electrons was calculated by subtracting the contribution of phonon from the measured heat capacity. Using the alpha model [28], the normalized specific heat jump can be estimated: $\Delta C_{el}/\gamma T_c = 2.89$ for $x = 0.014$ (seen in Figure 3(e)), which can reflect the strength of coupling. According to BCS theory, the ratio $\Delta C_{el}/\gamma T_c$ of weak coupling is 1.43, which is much smaller than the value of the $x = 0.014$ sample, indicating the existence of strong coupling. In addition, the normalized specific heat jump shows a negative correlation trend with Ni content, demonstrating that the doping of Ni may weaken the electron-phonon coupling slightly, although all of them represent the behavior of strong electron-phonon coupling (as shown in Figure 3 and Table 1).

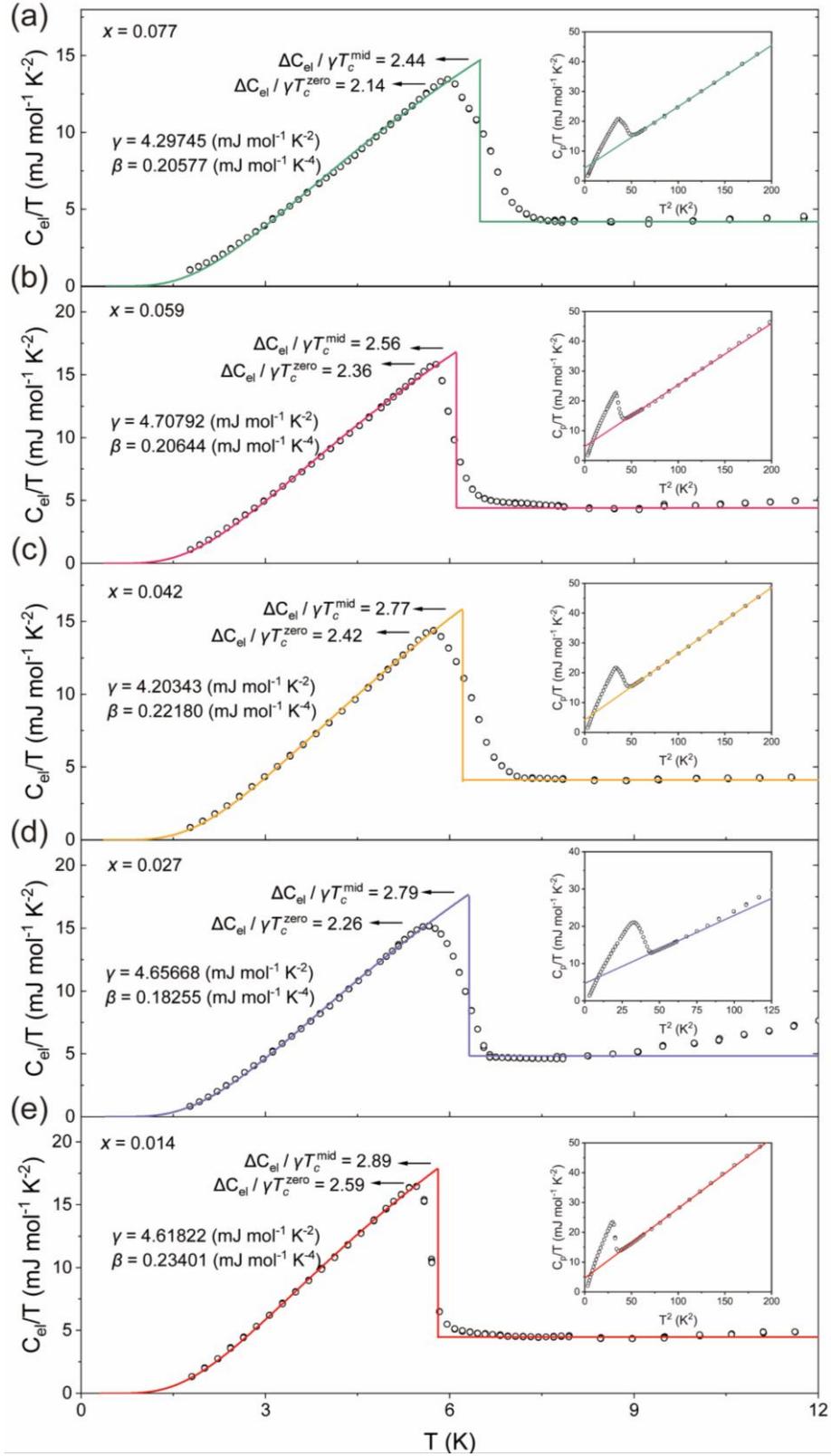

**Figure 3.** The measured electronic specific heat $C_{el}/T$ versus T for (a) $x = 0.077$, (b) $x = 0.059$, (c) $x = 0.042$, (d) $x = 0.027$, (e) $x = 0.014$. The insets show the measured heat capacity $C_p/T$ versus $T^2$.

In addition, thermodynamic properties can be calculated through heat capacity measurements. The Debye temperature is calculated with $\Theta_D = \left(\frac{12\pi^4 nR}{5\beta}\right)^{1/3}$, where $n$ is the number of atoms per formula unit, and $R$ represents the gas constant. It is given $\Theta_D$ = 204 K for $x$ = 0.014. Using the McMillan formula: $\lambda_{ep} = \frac{1.04 + \mu^* \ln\left(\frac{\Theta_D}{1.45 T_c}\right)}{(1 - 0.62\mu^*) \ln\left(\frac{\Theta_D}{1.45 T_c}\right) - 1.04}$, where $\mu^*$ is a Coulomb pseudo-potential and has a typical value of 0.13 for intermetallic superconductor [29,30], the electron-phonon coupling constant $\lambda_{ep}$ is estimated as 0.81, which can be classified as strongly-coupled superconductor. The analysis of these thermodynamic parameters provides profound insights into the fundamental mechanisms underlying the superconducting behavior observed in the sample series. The logarithmically averaged characteristic phonon frequency $\omega_{ln}$ can be calculated by the formula $\frac{\Delta C_{el}}{\gamma T_c} = 1.43[1 + 53(\frac{T_c}{\omega_{log}})^2 \ln(\frac{\omega_{log}}{3T_c})]$. The relationship between normalized specific heat jump and $\omega_{ln}$ normalized by $T_c$ is performed in Figure 4(a). One can observe that the specific heat jump decreases with the enhancement of normalized $\omega_{ln}$, following a similar trend with $(Ti_{1/3}Hf_{1/3}Ta_{1/3})_{1-x}Nb_x$ [31]. All of them are above the BCS weak coupling limit. According to the so-called alpha model, the superconducting gap values as well as the coupling strength $2\Delta_0/k_B T_c$ can be obtained by the equation $\Delta(T) = \alpha/\alpha_{BCS}\Delta_{BCS}(T)$, where $\alpha_{BCS}$ = 1.76 represents the ratio of weak-coupling. All these parameters are summarized in Table 1. As shown in Figure 4(b), the series of samples also follow a similar trend with $(Ti_{1/3}Hf_{1/3}Ta_{1/3})_{1-x}Nb_x$ and other strong coupling superconductors. The trend of $T_c$ can be interpreted by the McMillan equation modified by Allen and Dynes: $T_c = \frac{\omega_{ln}}{1.2} \exp[\frac{1.04(1+\lambda_{ep})}{\mu^*(1+0.62\lambda_{ep}) - \lambda_{ep}}]$. One can easily conclude that a high-$T_c$ superconductor should be high-$\omega_{ln}$. The $\omega_{ln}$ of our samples show a positive correlation with content $x$, which is demonstrated in Figure 4(c) and Table 1. These phenomena may result in the enhancement of $T_c$. The heat capacity measurements have proved that our samples are strong coupling superconductors. Therefore, the electron-phonon coupling constant $\lambda_{ep}$ can be calculated by the modified McMillan equation mentioned above, which is shown in Figure S4. The $\lambda_{ep}$ decreases with the enhancement of the content $x$, consistent with the decreases in normalized specific heat jump.

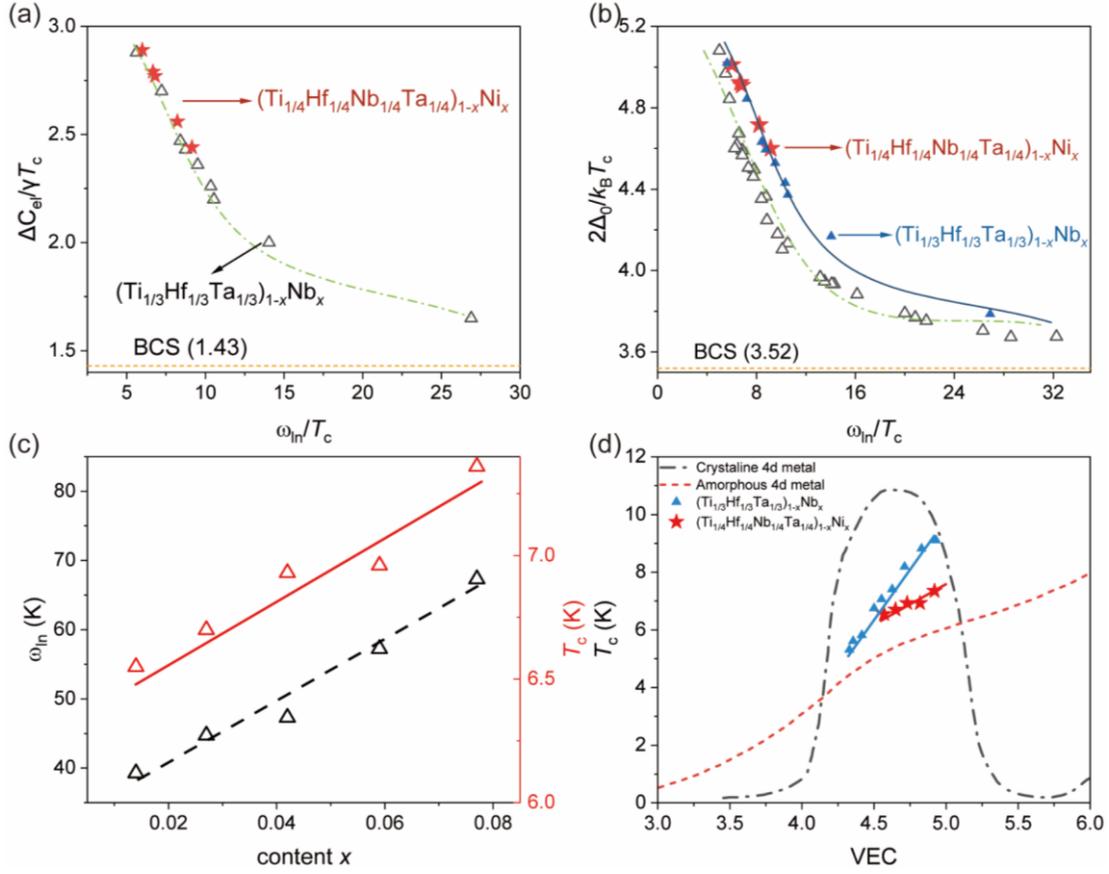

**Figure 4.** (a) The relationship between normalized specific heat jump $\Delta C_{el}/\gamma T_c$ and $\omega_{ln}/T_c$. (b) The coupling strength $2\Delta_0/k_B T_c$ VS the average phonon frequency $\omega_{ln}$ normalized by $T_c$, the white triangle represents other superconductors (refs. [23,32-35]). (c) The average phonon frequency $\omega_{ln}$ and $T_c$ as a function of the VEC. (d) The VEC phase diagram of different kinds of superconductors (refs. [14, 31, 35-37]).

Studies have revealed that the valence electron count (VEC) of the MEA or HEA superconductors plays a crucial role in determining their $T_c$ values. The VEC-dependence $T_c$s of HEA follows a similar trend as Ni content. To investigate Ni's contribution to superconductivity, we plot the relationship between $T_c$ and VEC of different superconductors in Figure 4(d). As shown in Figure 4(d), the trend of $T_c$s for crystalline transition metals, which is known as the Matthias rule, rises monotonically with the enhancement of VEC until its maximum appears at VEC = 4.7. The tendency of $T_c$s for amorphous transition metal vapor-deposited films is also displayed in Figure 4(d), which increases monotonically without reaching a maximum. The observed increase in $T_c$ in $(Ti_{1/4}Hf_{1/4}Nb_{1/4}Ta_{1/4})_{1-x}Ni_x$ is relatively similar to that in crystalline alloys. Besides, the correlation between $T_c$ and VEC in $(Ti_{1/4}Hf_{1/4}Nb_{1/4}Ta_{1/4})_{1-x}Ni_x$ is highly consistent with that observed in $(Ti_{1/3}Hf_{1/3}Ta_{1/3})_{1-x}Nb_x$ and $(TaNb)_{1-x}(ZrHfTi)_x$ [14, 31]. Therefore, it's evident that the enhancement of $T_c$ may come from the increase of VEC. We can conclude that the doping of Ni increases the VEC of our system, resulting in the enhancement of $T_c$. This unique phenomenon may provide a platform for investigating the superconductivity between crystalline and amorphous metals.

In low temperatures, the resistivity can be written by this equation: $\rho(T) = \rho_0 + AT^2$ as well as the electron-specific heat $C_{el}(T) = \gamma T$. The quadratic term of resistivity $AT^2$ represents the contribution of electron-electron scattering. Also, the Sommerfeld constant is connected with the effective mass of the electron. Therefore, the Kadowaki-Woods ratio (KWR), KWR = $A/\gamma^2$ is used to represent the correlation strength of the electron [37-40]. Research has found that the KWR of an amount of transition metals is a constant, which is $A/\gamma^2 \approx a_{TM} = 0.4$ μΩ cm mol$^2$ K$^2$ J$^{-2}$ = $0.04a_0$, where $a_0$ is a constant ($a_0 = 10$ μΩ cm mol$^2$ K$^2$ J$^{-2}$). The same phenomena happen in heavy fermion compounds, which is $A/\gamma^2 \approx a_{HF} = 10$ μΩ cm mol$^2$ K$^2$ J$^{-2}$ = $a_0$. The coefficient A of $(Ti_{1/4}Hf_{1/4}Nb_{1/4}Ta_{1/4})_{0.923}Ni_{0.077}$ is estimated as A = $1.5 \times 10^{-3}$ μΩ cm K$^{-2}$, while the Sommerfeld constant $\gamma = 4.297$ mJ mol$^{-1}$ K$^{-2}$. Figure 5 displayed the coefficient A vs the Sommerfield coefficient $\gamma$ for various compounds. The $(Ti_{1/4}Hf_{1/4}Nb_{1/4}Ta_{1/4})_{1-x}Ni_x$ is located beyond the line of heavy fermion, indicating the appearance of extremely strong electron correlation compared with other HEA. This also means that the doping of Ni enhances the strength of electron correlation of the Ti-Zr-Hf-Nb-Ta system HEAs, providing a new platform to study the interplay between $s$-wave superconductivity and strong electron correlation.

Various scenarios have been suggested to account for the surprisingly high KWR in other compounds. It depends on the effects of anisotropy, Fermi-surface topological, ground state degeneracy, magnetic correlation, etc [37,41,42]. Considering the characteristics of strong coupling in the $(Ti_{1/4}Hf_{1/4}Nb_{1/4}Ta_{1/4})_{1-x}Ni_x$ ($0 < x \leq 0.077$) case, we propose that the mechanism behind the larger KWR might be akin to that found in Nb$_3$Sn and V$_3$Si [37,43], which suggested that the strong dynamical coupling between conduction electrons and phonons may give rise to the heavy fermion bands at low temperatures. However, it is worth noting that our KWR value is nearly an order of magnitude larger than that obtained in V$_3$Si and many heavy fermion compounds, suggesting that additional mechanisms should be taken into account.

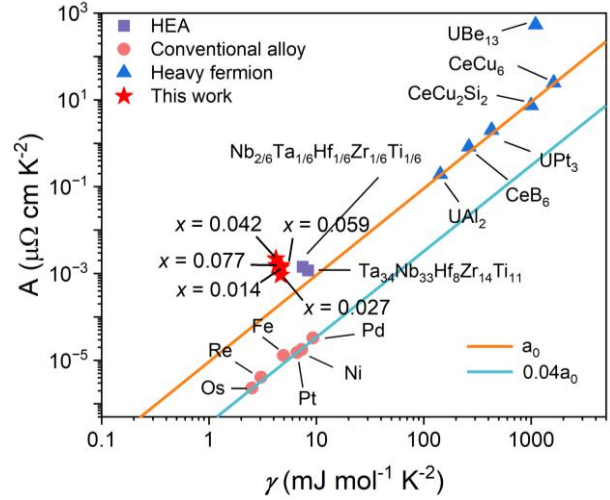

**Figure 5.** The coefficient A vs the Sommerfield coefficient $\gamma$ for various compounds (refs. [11, 32, 37-40]).

## IV. Conclusion

In conclusion, a series of MEA-HEA $(Ti_{1/4}Hf_{1/4}Nb_{1/4}Ta_{1/4})_{1-x}Ni_x$ ($0 < x \leq 0.077$) are synthesized by an arc-melted method, all of which show BCC structure. Resistivity, magnetic susceptibility, and specific heat measurement results indicate all $(Ti_{1/4}Hf_{1/4}Nb_{1/4}Ta_{1/4})_{1-x}Ni_x$ samples are type-II superconductors. $T_c$s increase with the increase of Ni-doping content. The specific heat jump ($\Delta C_{el}/\gamma T_c^{mid}$) of all $(Ti_{1/4}Hf_{1/4}Nb_{1/4}Ta_{1/4})_{1-x}Ni_x$ are much larger than the BCS value of 1.43, suggesting all these HEAs are strongly coupled superconductors. Additionally, large KWR values indicate that there is a strong electron correlation effect in this system. Our findings on $(Ti_{1/4}Hf_{1/4}Nb_{1/4}Ta_{1/4})_{1-x}Ni_x$ shed new light on the exploration of potential unconventional superconductivity in HEAs with 3$d$-transition metal elements with strong magnetism. The $(Ti_{1/4}Hf_{1/4}Nb_{1/4}Ta_{1/4})_{1-x}Ni_x$ HEA system is a new ideal material platform for the study of strong correlation behavior and strongly coupled superconductivity, which provides an insight into the physics of high-temperature superconductors or other unconventional superconductors.

## Acknowledgments

Acknowledgments

This work is funded by the Natural Science Foundation of China (No. 12274471, 12404165, 11922415), Guangdong Basic and Applied Basic Research Foundation (No. 2025A1515010311), Guangzhou Science and Technology Programmer (No. 2024A04J6415), the State Key Laboratory of Optoelectronic Materials and Technologies (Sun Yat-Sen University, No. OEMT-2024-ZRC-02), the Fundamental Research Funds for the Central Universities, Sun Yat-sen University (24qnpy092), Key Laboratory of Magnetoelectric Physics and Devices of Guangdong Province (Grant No. 2022B1212010008), and Research Center for Magnetoelectric Physics of Guangdong Province (2024B0303390001). Lingyong Zeng acknowledges the Postdoctoral Fellowship Program of CPSF (GZC20233299) and the International Postdoctoral Exchange Fellowship Program 2024 of the Human Resources and Social Security Department of Guangdong Province.